\documentclass{article}
\usepackage{spconf,amsmath,graphicx}
\usepackage{booktabs}
\usepackage{amssymb}
\usepackage{amsfonts}
\usepackage{multirow}
\usepackage[inline]{enumitem}
\usepackage{cleveref}

\title{Investigating Local and Global Information for Automated Audio Captioning with Transfer Learning}
\name{Xuenan Xu,
      Heinrich Dinkel,
      Mengyue Wu,
      Zeyu Xie,
      Kai Yu\thanks{Mengyue Wu and Kai Yu are the corresponding authors.}}
\address{MoE Key Lab of Artificial Intelligence\\
SpeechLab, Department of Computer Science and Engineering\\
AI Institute, Shanghai Jiao Tong University, Shanghai, China\\
        \{\textit{wsntxxn, richman, mengyuewu, kai.yu}\}@sjtu.edu.cn, zeyuxie29@gmail.com\\ 
 }
\begin{document}
%
\maketitle
\begin{abstract}
Automated audio captioning (AAC) aims at generating summarizing descriptions for audio clips. 
Multitudinous concepts are described in an audio caption, ranging from local information such as sound events to global information like acoustic scenery.
Currently, the mainstream paradigm for AAC is the end-to-end encoder-decoder architecture, expecting the encoder to learn all levels of concepts embedded in the audio automatically. 
This paper first proposes a topic model for audio descriptions, comprehensively analyzing the hierarchical audio topics that are commonly covered.
We then explore a transfer learning scheme to access local and global information.
Two source tasks are identified to respectively represent local and global information, being Audio Tagging (AT) and Acoustic Scene Classification (ASC).
Experiments are conducted on the AAC benchmark dataset Clotho and Audiocaps, amounting to a vast increase in all eight metrics with topic transfer learning. 
Further, it is discovered that local information and abstract representation learning are more crucial to AAC than global information and temporal relationship learning.  
\end{abstract}
\section{Introduction}
\label{sec:intro}
Automated audio captioning (AAC) is a cross-modal task bridging audio signal processing and natural language processing  (NLP)~\cite{drossos2017automated,wu2019audio}.
The introduction of AAC to the Detection and Classification of Acoustic Scenes and Events (DCASE) 2020 challenge sparked interest from researchers~\cite{koizumi2020transformer,wu2020audio,xu2020crnn,cakir2020multi}.
AAC is particularly interesting yet challenging as the audio captions describe multitudinous auditory elements. 
Compared with visual perception, where the objects are defined by its shape, color, size, and its spatial position to other objects, auditory perception concerns with sound events and their corresponding physical properties, temporal information of these sound events, and their relationship with other events, and high-level knowledge-rich auditory understanding. 
For instance, a typical caption from the DCASE benchmark dataset Clotho~\cite{drossos2020clotho} ``people talking in a small and empty room'' describes the sound event ``people talking'' and its global scene ``in a room'', where high-level auditory knowledge is processed to infer that the room is small and empty, a visual description.

\begin{figure*}[!t]
    \centering
    \includegraphics[width=0.95\linewidth]{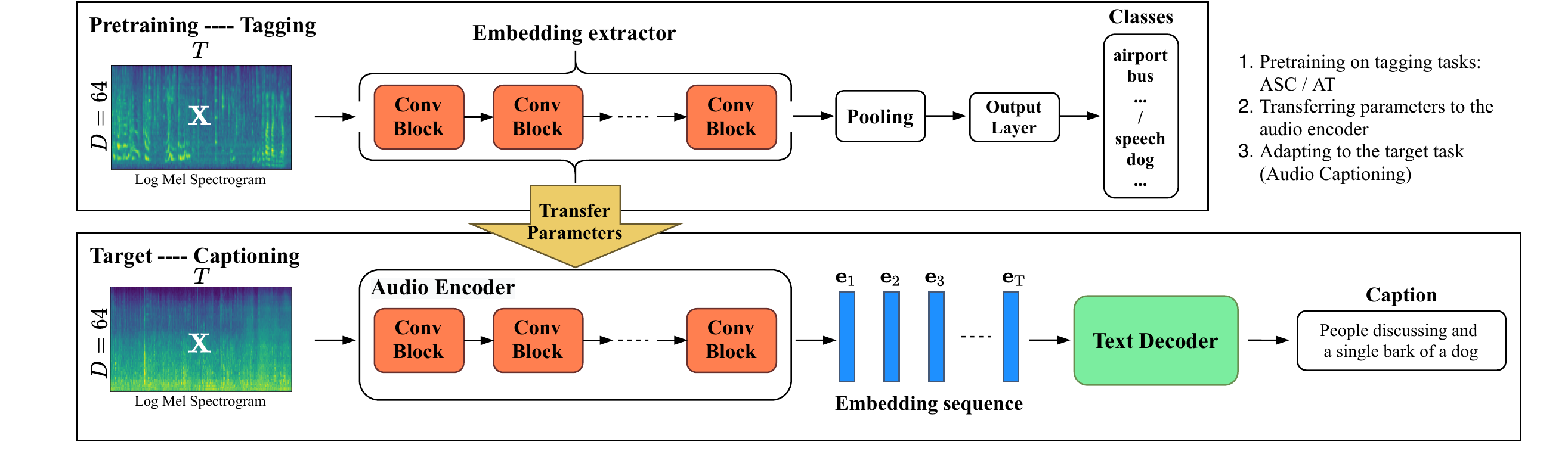}
    \caption{Our proposed transfer learning for automated audio captioning. 
    In the first stage, a tagging system is pretrained by ASC or AT. 
    Then the embedding extractor part of the pretrained tagging system is used to initialize the audio encoder. 
    In the second stage, the audio encoder is adapted to the target task (AAC), and the entire captioning system is trained end-to-end.}
    \label{fig:pretrain_scheme}
\end{figure*}

It should be noted that the current mainstream training paradigm of AAC is the end-to-end encoder-decoder framework, where the captions are provided as the only supervision signal to the audio content. 
An audio encoder first extracts an abstract embedding from an input audio clip in the encoder-decoder framework. 
Then the text decoder predicts the caption according to the audio embedding. 
To encode all the above-mentioned multifaceted information from an audio clip without explicit supervision increases the difficulty for AAC encoder training. 
Therefore, a hierarchical structure of the abstract audio topics commonly described in audio captions is crucial. 

Stemming from auditory perception and combined with the captions provided in the currently available AAC datasets, we propose the following audio topic model for AAC:
\begin{enumerate}
    \item Local audio topics:
    \begin{enumerate*}
        \item Sound events, which can be described by the sounding object entity (``a male''), the verbs that make the sound (``talk''), the physical properties of the sound (``loud''). 
    \end{enumerate*}
    \item Global audio topics:
    \begin{enumerate*}
        \item Acoustic scenes, such as an exact scene location description (``downtown''), and an abstract description (``in the distance'').
        \item High-level abstraction, including content inference (``at a conference''), and affect expression (``annoyingly'').
    \end{enumerate*}
\end{enumerate}

We explore a transfer learning method to address local and global information based on such a topic model.
Two source tasks are identified to represent local and global information, being Audio Tagging (AT) and Acoustic Scene Classification (ASC).
ASC is an environmental sound recognition task, attempting to classify global audio representations into predefined scene categories, e.g., park, shopping mall.
On the other hand, AT aims at identifying specific sound events present in an audio recording.
We propose pretraining the audio encoder on ASC and AT tasks and then transferring the parameters to the AAC encoder, as shown in \Cref{sec:tagging_pretrain}. 
Since AAC concerns both abstract representations and the sound events' temporal relationships, different pretraining backbone networks are explored.
Experiments in \Cref{sec:exp_setup} are conducted on the benchmark AAC dataset Clotho and Audiocaps, the largest AAC dataset by far. 
A consistent performance gain is obtained over the eight language similarity metrics on both datasets, displayed in \Cref{sec:res_discuss}.

\section{Transfer learning for AAC}
\label{sec:tagging_pretrain}
We propose a transfer learning approach for a more effective AAC audio encoder. 
For the following definitions, assume that $\mathbf{X} \in \mathbb{R}^{T\times D}$ is an input feature with $T$ frames and $D$ Mel filters.
The supervised pretraining tasks used in this paper are modeled as $\mathcal{F}: \mathbf{X} \mapsto y$, where $y$ differs for each respective task.
For our research, we experiment with two different source tasks, being audio tagging (AT) $\mathcal{F}_{tag}: \mathbf{X} \mapsto \mathbf{y}_{\text{tag}}, \mathbf{y}_{\text{tag}} \in \{0,1\}^{E}$ and acoustic scene classification (ASC) $\mathcal{F}_{asc}: \mathbf{X} \mapsto y_{\text{asc}}, y_{\text{asc}} \in \{1,\ldots,E\}$.

The process of our proposed transfer learning for AAC is illustrated in \Cref{fig:pretrain_scheme}.
The backbone encoder architecture comprises an embedding extractor, followed by a temporal pooling layer and an output layer.
The embedding extractor consists of several convolution blocks to extract mid-level embedding features from the audio input.
After pretraining, the parameters of the AT / ASC system are transferred to the AAC audio encoder.
We experiment with a CNN and a CRNN pretraining encoder network on both the AT and ASC tasks. 
We intend to explore whether abstract embeddings (CNN) or temporal information (CRNN) have a more significant impact on AAC performance.

\textbf{AAC Model Architecture}
\begin{figure*}[!htpb]
    \centering
    \includegraphics[width=\textwidth]{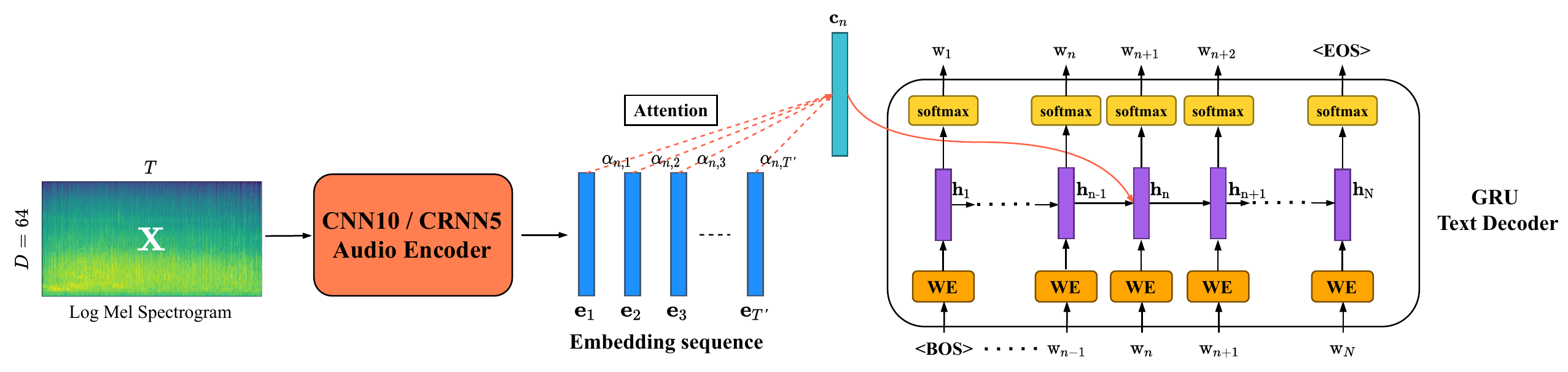}
    \caption{The illustration of the encoder-decoder AAC system. The audio encoder outputs a sequence of feature embeddings $\mathbf{e}_1, \mathbf{e}_2, \cdots, \mathbf{e}_{T^{\star}}$ from a Log Mel Spectrogram input with $T$ frames. The attentional GRU text decoder updates its hidden state $\mathbf{h}_n$ based on the previous hidden $\mathbf{h}_{n-1}$, the current word $\text{w}_n$ and a context vector $\mathbf{c}_n$, which is a weighted combination of the embedding sequence.}
    \label{fig:model_architecture}
\end{figure*}
Regarding the audio caption framework, we adopt a temporal attentional encoder-decoder architecture for AAC.
The overview of our system is presented in \Cref{fig:model_architecture}.
It consists of an audio encoder which extracts feature embedding sequence from a Log Mel Spectrogram (LMS) input and a text decoder to output caption.
The whole system is trained by the standard cross-entropy loss between the predicted caption and ground truth annotation.

\textbf{Audio encoder}
As mentioned previously, two model architectures are adopted as the audio encoder for comparison purposes: a 10-layer CNN (CNN10) and a 5-layer CRNN (CRNN5).
CNN10 shows superior performance for time-invariant audio tagging~\cite{kong2020panns}, while CRNN5 is reported to be effective in duration robust sound event detection~\cite{dinkel2020duration}.
For either architecture, the audio encoder reads an LMS input with $T$ frames and outputs a sequence of embeddings $\{\mathbf{e}_t\}_{t=1}^{T^{\star}}$ ($T^{\star}$ may not equal $T$ due to temporal subsampling).
The detailed structure of CNN10 and CRNN5 can be found in~\cite{kong2020panns} and~\cite{dinkel2020duration}.

\textbf{Text decoder}
We adopt a standard shallow single layer unidirectional GRU as the text decoder.
At decoding timestep $n$, the hidden state $\mathbf{h}_n$ is updated depending on the previous timestep $\mathbf{h}_{n-1}$, the current input word $\text{w}_n$ and a context vector $\mathbf{c}_n$ as:
$$
    \mathbf{h}_n = \text{GRU}([\mathbf{c}_n; \text{WE}(\text{w}_n)], \mathbf{h}_{n-1})
$$
where WE denotes the word embedding layer.
A standard temporal attention mechanism~\cite{bahdanau2015neural} is utilized to obtain $\mathbf{c}_n$.
The attention weights $\{\alpha_{n,t}\}_{t=1}^{T^{\star}}$ are calculated by aligning $\mathbf{h}_{n-1}$ with the timesteps of audio embeddings:
\begin{align}
    \alpha_{n,t} &= \frac{\exp\big(\text{score}(\mathbf{h}_{n-1}, \mathbf{e}_t)\big)}{\sum_{t=1}^{T^{\star}} \exp\big(\text{score}(\mathbf{h}_{n-1}, \mathbf{e}_t)\big)}\\
    \mathbf{c}_n &= \sum_{t=1}^{T^{\star}}\alpha_{n,t}\mathbf{e}_t
\end{align}
We use the \textit{concat} scoring function~\cite{luong2015effective} for the alignment.
After a fully connected output layer and a softmax function, the output word probabilities are obtained.
The word with the highest probability is selected as the output.
This process is repeated until an end-of-sentence (EOS) token is reached.

\section{Experimental setup}
\label{sec:exp_setup}

\subsection{Datasets}
Regarding the pretraining target data source, we use AudioSet for AT and DCASE for ASC.
For the target AAC task, we conduct our experiments on Clotho dataset~\cite{drossos2020clotho} and Audiocaps~\cite{kim2019audiocaps}.

\textbf{AudioSet}~\cite{gemmeke2017audio} is a large-scale manually-annotated sound event dataset.
Each audio clip has a duration of up to ten seconds, containing at least one sound event label.
AudioSet consists of a 527 event ontology, encompassing most everyday sounds.
Two training sets are provided: a balanced (60 hours, 20k+ clips) and an unbalanced (5000 hours, 1.85 million clips) one.

\textbf{DCASE}  We incorporate the development sets from ASC task (Task1A) of DCASE2019 and DCASE2020 challenges.
DCASE2019 contains 26 hours of audio recorded in urban scene environments, and DCASE2020 comprises 39 hours, totaling 65 hours of data.
To avoid a bias due to different data sizes, we randomly selected 60 hours of data from those two datasets, matching the balanced AudioSet size.

\begin{table*}[!htpb]
    \centering
    \caption{Performance on the Clotho and Audiocaps evaluation set of different pretraining settings. KWP, ASC, AT denote keyword prediction, acoustic scene classification and audio tagging respectively. KWP
    achieves the best single model performance on Clotho evaluation set, while VGG features is the best AudioCaps model. $\text{B}_{\text{N}}$ represents the N-gram BLEU score.}
    \begin{tabular}{rrr||lllllllll}
    \toprule
    Data & Encoder & Pretrain & $\text{B}_1$ & $\text{B}_2$ &$\text{B}_3$ & $\text{B}_4$ & $\text{ROUGE}_\text{L}$ & CIDEr & METEOR & SPICE\\
    \midrule
    \parbox[t]{2.5mm}{\multirow{9}{*}{\rotatebox[origin=c]{90}{Clotho}}}& CNN10 & KWP~\cite{wu2020audio} & 53.4 & 34.3 & 23.0 & 15.1 & 35.6 & 34.6 & 16.0 & 10.8\\
    \cline{2-11}
     & \multirow{4}{*}{CNN10} & from scratch & 47.5 & 27.5 & 17.6 & 11.1 & 31.9 & 21.3 & 13.6 & 8.1 \\
     &  & ASC & 50.7 & 30.9 & 20.1 & 12.7 & 33.9 & 27.5 & 14.9 & 9.2\\
     &  & AT (balanced) & 53.1 & 33.1 & 21.6 & 13.9 & 35.5 & 31.9 & 16.0 & 10.4\\
     &  & AT (unbalanced) & \textbf{55.6} & \textbf{36.3} & \textbf{24.2} & \textbf{15.9} & \textbf{36.8} & \textbf{37.7} & \textbf{16.9} & \textbf{11.5}\\
    \cline{2-11}
     & \multirow{4}{*}{CRNN5} & from scratch & 51.0 & 31.4 & 20.6 & 13.5 & 34.4 & 28.5 & 15.0 & 9.3 \\
     &  & ASC & 49.5 & 30.1 & 19.3 & 12.0 & 33.4 & 25.7 & 14.5 & 9.1\\
     &  & AT (balanced) & 52.3 & 32.6 & 21.4 & 13.8 & 34.9 & 29.9 & 15.7 & 10.2\\
     &  & AT (unbalanced) & 52.8 & 33.1 & 21.7 & 13.8 & 35.2 & 30.1 & 15.6 & 10.1\\
    \hline
    \midrule
    \parbox[t]{2.5mm}{\multirow{9}{*}{\rotatebox[origin=c]{90}{AudioCaps}}} & Multiscale & VGG features~\cite{kim2019audiocaps} & 61.4 & 44.6 & 31.7 & 21.9 & 45.0 & 59.3 & 20.3 & 14.4 \\ 
    \cline{2-11}
     & \multirow{4}{*}{CNN10} & from scratch & 62.1 & 44.0 & 30.3 & 20.5 & 44.2 & 52.1 & 20.6 & 14.1\\
     & & ASC & 62.7 & 44.2 & 30.5 & 20.7 & 43.9 & 54.1 & 20.5 & 14.5\\
     & & AT (balanced) & 63.8 & 46.1 & 32.3 & 22.0 & 45.1 & 57.8 & 21.5 & 15.3\\
     &  & AT (unbalanced) & \textbf{65.5} & \textbf{47.6} & \textbf{33.5} & \textbf{23.1} & \textbf{46.7} & \textbf{66.0} & \textbf{22.9} & \textbf{16.8} \\
    \cline{2-11}
     & \multirow{4}{*}{CRNN5} & from scratch & 61.9 & 44.5 & 31.1 & 21.0 & 44.6 & 54.5 & 20.8 & 14.6\\
     & & ASC & 60.5 & 43.2 & 30.2 & 21.0 & 43.3 & 51.5 & 19.8 & 14.1\\
     & & AT (balanced) & 62.9 & 45.4 & 32.1 & 22.6 & 45.0 & 60.2 & 20.7 & 14.9\\
     & & AT (unbalanced) & 64.1 & 46.6 & 33.2 & 22.8 & 46.0 & 60.5 & 21.5 & 15.9\\ 
    \bottomrule
    \end{tabular}
    \label{tab:result}
\end{table*}

\textbf{Clotho}~\cite{drossos2020clotho} is a newly published AAC benchmark dataset used for the DCASE2020 task 6 challenge.
There are 2893 audio clips (18 hours) in the development set and 1043 clips (7 hours) in the evaluation set, ranging evenly from 15 to 30 seconds in duration.
Each audio clip has five corresponding caption annotations.

\textbf{Audiocaps}~\cite{kim2019audiocaps} is by far the largest AAC dataset, consisting of 46k audio clips ($\approx$ 127 hours) collected from the AudioSet dataset. 
One human-annotated caption is provided for the training dataset while five captions for validation and test sets, respectively. 

\vspace{-3mm}
\subsection{System configuration}

Standard 64-dimensional LMS features are extracted every 20 ms with a Hann window size of 40 ms.
For both pretraining and AAC fine-tuning, 90\% of the development set is split as the training subset, and the rest 10\% is used for cross-validation.
The initial learning rates are set to $\mathrm{10}^{-3}$ and $\mathrm{5}\times10^{-4}$, batch sizes to 64 and 32 for AT / ASC pretraining and AAC fine-tuning, respectively.
During pretraining, early stopping is utilized, where the model with the best performance on the validation set is chosen for encoder initialization.
Training is uniformly done using Adam optimization~\cite{kingma2014adam}.
The standard caption metrics (BLEU@1-4~\cite{papineni2002bleu}, ROUGE~\cite{lin2004rouge}, METEOR~\cite{banerjee2005meteor}, CIDEr~\cite{vedantam2015cider}, and SPICE~\cite{anderson2016spice}) are used for evaluation.
Beam search with a beam size of 3 is adopted during evaluation to enhance performance.

\section{Results and discussion}
\label{sec:res_discuss}
\Cref{tab:result} presents our results on Clotho and Audiocaps datasets, with and without encoder pretraining, respectively elaborating local (AT) and global (ASC) information. 
The majority of the pretraining approaches, except for CRNN5 pretraining on the ASC task, enhance the performance compared with training from scratch.
This indicates that a captioning model embeds different information levels and encoder pretraining on relevant tasks can help the model attend to further details.
We further compare our approach against the best-published results, i.e., an encoder pretraining method by keyword prediction (KWP)~\cite{wu2020audio} achieves the best \textbf{single} model performance on Clotho evaluation set. 
In contrast, a multi-scale encoder with pretrained AudioSet VGGish features~\cite{kim2019audiocaps} is the state-of-the-art (SOTA) performance on Audiocaps with only audio inputs.
Our CNN10 encoder pretraining on AT (unbalanced) achieves the best result on Clotho and Audiocaps, outperforming previous work.
\vspace{-4mm}
\paragraph*{Local vs. Global}
We deliberately choose AT and ASC tasks to represent local and global audio topics in AAC, corresponding to the two pretraining tasks' characteristics: AT provides detailed audio event information, while ASC aims to characterize the environment.
Results on both Clotho and Audiocaps indicate that local audio topics are comparatively more crucial to a captioning model than global information: AAC with AT pretraining always outperform ASC pretraining. 
In particular, AT pretraining on unbalanced constantly yields the best performance, amounting to SOTA performance on both datasets, regardless of the evaluation metrics.
Even when AT (balanced) and ASC datasets contain approximately the same amount of data ($\approx$ 60 h), AAC performance significantly improves when pretraining on AT.
\vspace{-4mm}
\paragraph*{Abstract embedding vs. Temporal information}
In addition to local vs global information, we explore whether abstract embeddings (CNN) or temporal information (CRNN) have a more significant impact.
When trained from scratch, CRNN5 outperforms CNN10 on all metrics; in contrast to pretraining, CRNN5 brings little improvement for AAC performance.
This indicates that AAC prefers CNN10, which can better recognize the presence of audio events, focusing less on the temporal relationship of different events.
Performance improves steadily for all CNN10 models when training with more data (ASC, balanced, unbalanced).
Transferring knowledge learned via large dataset pretraining (AT unbalanced vs. balanced) improves the downstream AAC performance significantly, as in other work~\cite{kong2020panns,mun2017deep}.

\section{Conclusion}
This work investigates concepts commonly described in audio captions, referred to as an audio topic model.
Based on this, a transfer learning scheme is proposed to address the local and global information.
We compare two pretraining tasks (ASC and AT) and two audio encoder architectures (CNN10 and CRNN5) to investigate which abstract topic and architecture are crucial to AAC.  
The results show that transferring knowledge from either topic leads to vast performance gain, leading to SOTA performance on both Clotho and Audiocaps.
It is observed that local information (AT) and abstract embeddings (CNN10) are more critical to ACC.
We would like to explore methods like multi-task training to better address the different topics within a caption for future work. 
Topic fusion could also shift from coarse to fine-scale, e.g., separately modeling different traits of sound events, relationships, exact and abstract acoustic scenes, along with the high-level knowledge-infused abstraction.   

\vspace{-0.2cm}
\section{Acknowledgement}
This work has been supported by National Natural Science Foundation of China (No.61901265), Shanghai Pujiang Program (No.19PJ1406300), and Startup Fund for Youngman Research at SJTU (No.19X100040009). Experiments have been carried out on the PI supercomputer at Shanghai Jiao Tong University.
\vfill\pagebreak

\bibliographystyle{IEEEtran}
\bibliography{refs}

\begin{thebibliography}{10}
\providecommand{\url}[1]{#1}
\csname url@samestyle\endcsname
\providecommand{\newblock}{\relax}
\providecommand{\bibinfo}[2]{#2}
\providecommand{\BIBentrySTDinterwordspacing}{\spaceskip=0pt\relax}
\providecommand{\BIBentryALTinterwordstretchfactor}{4}
\providecommand{\BIBentryALTinterwordspacing}{\spaceskip=\fontdimen2\font plus
\BIBentryALTinterwordstretchfactor\fontdimen3\font minus
  \fontdimen4\font\relax}
\providecommand{\BIBforeignlanguage}[2]{{%
\expandafter\ifx\csname l@#1\endcsname\relax
\typeout{** WARNING: IEEEtran.bst: No hyphenation pattern has been}%
\typeout{** loaded for the language `#1'. Using the pattern for}%
\typeout{** the default language instead.}%
\else
\language=\csname l@#1\endcsname
\fi
#2}}
\providecommand{\BIBdecl}{\relax}
\BIBdecl

\bibitem{drossos2017automated}
K.~Drossos, S.~Adavanne, and T.~Virtanen, ``Automated audio captioning with
  recurrent neural networks,'' in \emph{IEEE Workshop on Applications of Signal
  Processing to Audio and Acoustics (WASPAA)}.\hskip 1em plus 0.5em minus
  0.4em\relax IEEE, 2017, pp. 374--378.

\bibitem{wu2019audio}
M.~Wu, H.~Dinkel, and K.~Yu, ``Audio caption: Listen and tell,'' in \emph{Proc.
  IEEE International Conference on Acoustics, Speech and Signal Processing
  (ICASSP)}.\hskip 1em plus 0.5em minus 0.4em\relax IEEE, 2019, pp. 830--834.

\bibitem{koizumi2020transformer}
Y.~Koizumi, R.~Masumura, K.~Nishida, M.~Yasuda, and S.~Saito, ``A
  transformer-based audio captioning model with keyword estimation,'' in
  \emph{Proc. ISCA Interspeech}, 2020, pp. 1977--1981.

\bibitem{wu2020audio}
Y.~Wu, K.~Chen, Z.~Wang, X.~Zhang, F.~Nian, S.~Li, and X.~Shao, ``{Audio
  Captioning Based On Transformer And Pre-trained CNN},'' in \emph{Proc.
  Detection and Classification of Acoustic Scenes and Events Workshop (DCASE)},
  no. November, 2020, pp. 21--25.

\bibitem{xu2020crnn}
X.~Xu, H.~Dinkel, M.~Wu, and K.~Yu, ``A crnn-gru based reinforcement learning
  approach to audio captioning,'' in \emph{Proc. Detection and Classification
  of Acoustic Scenes and Events Workshop (DCASE)}, Tokyo, Japan, November 2020,
  pp. 225--229.

\bibitem{cakir2020multi}
E.~Çakır, K.~Drossos, and T.~Virtanen, ``Multi-task regularization based on
  infrequent classes for audio captioning,'' in \emph{Proc. Detection and
  Classification of Acoustic Scenes and Events Workshop (DCASE)}, Tokyo, Japan,
  November 2020, pp. 6--10.

\bibitem{drossos2020clotho}
K.~Drossos, S.~Lipping, and T.~Virtanen, ``Clotho: an audio captioning
  dataset,'' in \emph{Proc. IEEE International Conference on Acoustics, Speech
  and Signal Processing (ICASSP)}.\hskip 1em plus 0.5em minus 0.4em\relax IEEE,
  2020, pp. 736--740.

\bibitem{kong2020panns}
Q.~Kong, Y.~Cao, T.~Iqbal, Y.~Wang, W.~Wang, and M.~D. Plumbley, ``Panns:
  Large-scale pretrained audio neural networks for audio pattern recognition,''
  \emph{IEEE Trans. Audio, Speech, Language Process.}, vol.~28, pp. 2880--2894,
  2020.

\bibitem{dinkel2020duration}
H.~Dinkel and K.~Yu, ``Duration robust weakly supervised sound event
  detection,'' in \emph{Proc. IEEE International Conference on Acoustics,
  Speech and Signal Processing (ICASSP)}.\hskip 1em plus 0.5em minus
  0.4em\relax IEEE, 2020, pp. 311--315.

\bibitem{bahdanau2015neural}
D.~Bahdanau, K.~Cho, and Y.~Bengio, ``Neural machine translation by jointly
  learning to align and translate,'' in \emph{The International Conference on
  Learning Representations (ICLR)}, 2015.

\bibitem{luong2015effective}
M.-T. Luong, H.~Pham, and C.~D. Manning, ``Effective approaches to
  attention-based neural machine translation,'' in \emph{Proc. Conference on
  Empirical Methods in Natural Language Processing (EMNLP)}, 2015, pp.
  1412--1421.

\bibitem{kim2019audiocaps}
C.~D. Kim, B.~Kim, H.~Lee, and G.~Kim, ``Audiocaps: Generating captions for
  audios in the wild,'' in \emph{Proc. Conference of the North {A}merican
  Chapter of the Association for Computational Linguistics (NAACL)}, 2019, pp.
  119--132.

\bibitem{gemmeke2017audio}
J.~F. Gemmeke, D.~P. Ellis, D.~Freedman, A.~Jansen, W.~Lawrence, R.~C. Moore,
  M.~Plakal, and M.~Ritter, ``Audio set: An ontology and human-labeled dataset
  for audio events,'' in \emph{Proc. IEEE International Conference on
  Acoustics, Speech and Signal Processing (ICASSP)}.\hskip 1em plus 0.5em minus
  0.4em\relax IEEE, 2017, pp. 776--780.

\bibitem{kingma2014adam}
D.~P. Kingma and J.~Ba, ``Adam: {A} method for stochastic optimization,'' in
  \emph{The International Conference on Learning Representations (ICLR)}, 2015.

\bibitem{papineni2002bleu}
K.~Papineni, S.~Roukos, T.~Ward, and W.-J. Zhu, ``Bleu: a method for automatic
  evaluation of machine translation,'' in \emph{Proc. Annual Meeting of the
  Association for Computational Linguistics (ACL)}.\hskip 1em plus 0.5em minus
  0.4em\relax Association for Computational Linguistics, 2002, pp. 311--318.

\bibitem{lin2004rouge}
C.-Y. Lin, ``Rouge: A package for automatic evaluation of summaries,'' in
  \emph{Text summarization branches out}, 2004, pp. 74--81.

\bibitem{banerjee2005meteor}
S.~Banerjee and A.~Lavie, ``Meteor: An automatic metric for mt evaluation with
  improved correlation with human judgments,'' in \emph{Proc. acl workshop on
  intrinsic and extrinsic evaluation measures for machine translation and/or
  summarization}, 2005, pp. 65--72.

\bibitem{vedantam2015cider}
R.~Vedantam, C.~Lawrence~Zitnick, and D.~Parikh, ``Cider: Consensus-based image
  description evaluation,'' in \emph{Proc. IEEE Computer Society Conference on
  Computer Vision and Pattern Recognition (CVPR)}, 2015, pp. 4566--4575.

\bibitem{anderson2016spice}
P.~Anderson, B.~Fernando, M.~Johnson, and S.~Gould, ``Spice: Semantic
  propositional image caption evaluation,'' in \emph{European Conference on
  Computer Vision (ECCV)}.\hskip 1em plus 0.5em minus 0.4em\relax Springer,
  2016, pp. 382--398.

\bibitem{mun2017deep}
S.~Mun, S.~Shon, W.~Kim, D.~K. Han, and H.~Ko, ``Deep neural network based
  learning and transferring mid-level audio features for acoustic scene
  classification,'' in \emph{Proc. IEEE International Conference on Acoustics,
  Speech and Signal Processing (ICASSP)}.\hskip 1em plus 0.5em minus
  0.4em\relax IEEE, 2017, pp. 796--800.

\end{thebibliography}

\end{document}